\documentclass[preprint]{aastex}
\usepackage{graphicx}
\usepackage{url}
\usepackage{hyperref}
\makeatletter
\shorttitle{First Sunquake of Cycle 24 Observed by SDO}
\shortauthors{Kosovichev}
\bibpunct[, ]{(}{)}{; }{a}{,}{,}

\makeatother

\begin{document}

\title{First Sunquake of Solar Cycle 24 Observed by Solar Dynamics Observatory}

\author{A.G. Kosovichev}

\affil{ W.~W.~Hansen Experimental Physics Laboratory, Stanford University,
Stanford, CA 94305, USA }
\begin{abstract}
The X2.2-class solar flare of February 15, 2011, produced a powerful `sunquake' event, representing a seismic response to the flare impact. The impulsively excited seismic waves formed a compact wavepacket traveling through the solar interior and appeared on the surface as expanding wave ripples. The Helioseismic and Magnetic Imager (HMI), instrument on SDO, observes variations of intensity, magnetic field and plasma velocity (Dopplergrams) on the surface of Sun almost uninterruptedly with high resolution (0.5 arcsec/pixel) and high cadence (45 sec). The flare impact on the solar surface was observed in the form of compact and rapid variations of the HMI observables (Doppler velocity, line-of-sight magnetic field and continuum intensity). These variations, caused by the impact of high-energy particles in the photosphere, formed a typical two-ribbon flare structure. The sunquake can be easily seen in the raw Dopplergram differences without any special data processing. The source of this quake was located near the outer boundary of a very complicated complicated sunspot region, NOAA 1158, in a sunspot penumbra and at the penumbra boundary. This caused an interesting plasma dynamics in the impact region.  I present some preliminary results of analysis of the near-real-time data from HMI, and discuss  properties of the sunquake and the flare impact sources.

\end{abstract}


\section{Introduction}

``Sunquakes", the helioseismic response to solar flares, are
caused by strong localized hydrodynamic impacts in the photosphere
during the flare impulsive phase. The helioseismic waves are
observed directly as expanding circular-shaped ripples on the
solar surface, which can be detected in Dopplergram movies and as a
characteristic ridge in time-distance diagrams,
\citep{Kosovichev1998, Kosovichev2006a}, or  by
calculating integrated acoustic emission \citep{Donea1999,
Donea2005}.

Solar flares are sources of high-temperature plasma and
strong hydrodynamic motions in the solar atmosphere. Perhaps, in all
flares such perturbations generate acoustic waves traveling through
the interior. However, only in some flares is the impact
sufficiently localized and strong to produce the seismic waves with
the amplitude above the convection noise level. The sunquake events with
expanding ripples are relatively rare, and observed only  in
some high-M and  X-class flares. The last previous observation of the seismic waves
was during X1.2 flare of January 15, 2005.
\begin{figure}[t]
\begin{center}
\includegraphics[width=0.8\textwidth]{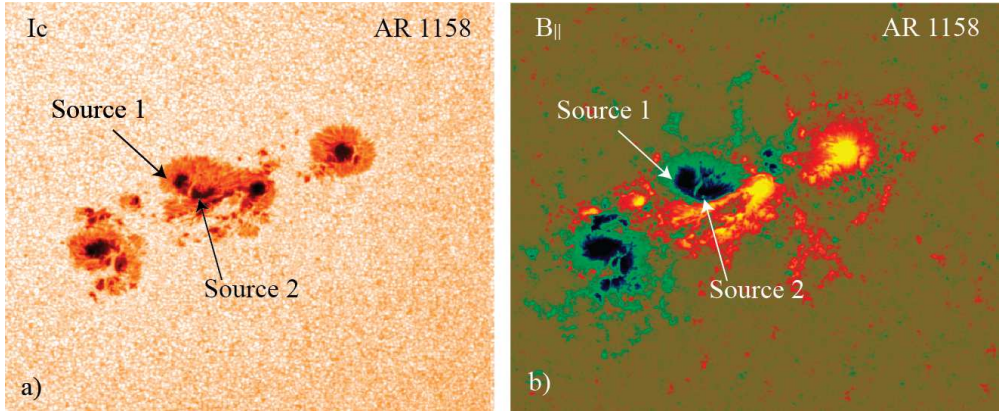}
\end{center}
\caption{a) Continuum intensity and b) line-of-site magnetic field images of AR 1158 during the X2.2 flare, at 2011.02.15\_01:54:39\_TAI. Arrows show positions of two analyzed sources of transient flare variations. A powerful sunquake originated from Source 1. Source 2 is a place of a strong impulsive impact, but it did not generate a significant sunquake.}\label{fig1}
\end{figure}

It has been found in the initial July 9, 1996, flare observations
\citep{Kosovichev1998} that the hydrodynamic impact follows the hard
X-ray flux impulse, and hence, the impact of high-energy electrons.
Thus, it was suggested that the mechanism of sunquakes can be explained
by a hydrodynamic thick-target model \citep{Kostiuk1975}. Several other
mechanisms, including impact by high-energy protons and back-warming
heating \citep{Donea2005}, and also due to magnetic field variations
\citep{Hudson2008}. However, the mechanism, which converts a part
of the flare energy and momentum into the seismic acoustic waves, is
currently unknown. It is also unknown why only some flares generate such waves.

Most of the previous observations of sunquakes were obtained with
the Michelson Doppler Imager instrument on SOHO. However, these
observations did not provide uninterrupted coverage. Generally,
the full-disk observations with the full 2 arcsec/pixel resolution
were obtained only for 2 months a year. Thus, many flares were not
observed, and the statistics of sunquakes and their relation to the
flare properties were not established.

Except short eclipse periods in March and September, the HMI
instrument  \citep{Schou2010} provides uninterrupted observations of the Sun,
and will help us to solve the mystery of `sunquakes'.
The flare of February 15, 2011, was the first X-class flare
of the new solar cycle, 24, and the first flare observed by
HMI. I present some preliminary results of initial analysis, which
reveal a sunquake event. This event had curious properties,
which make it different from the previously observed sunquakes.

\section{Results}
\begin{figure}[t]
\begin{center}
\includegraphics[width=0.3\textwidth]{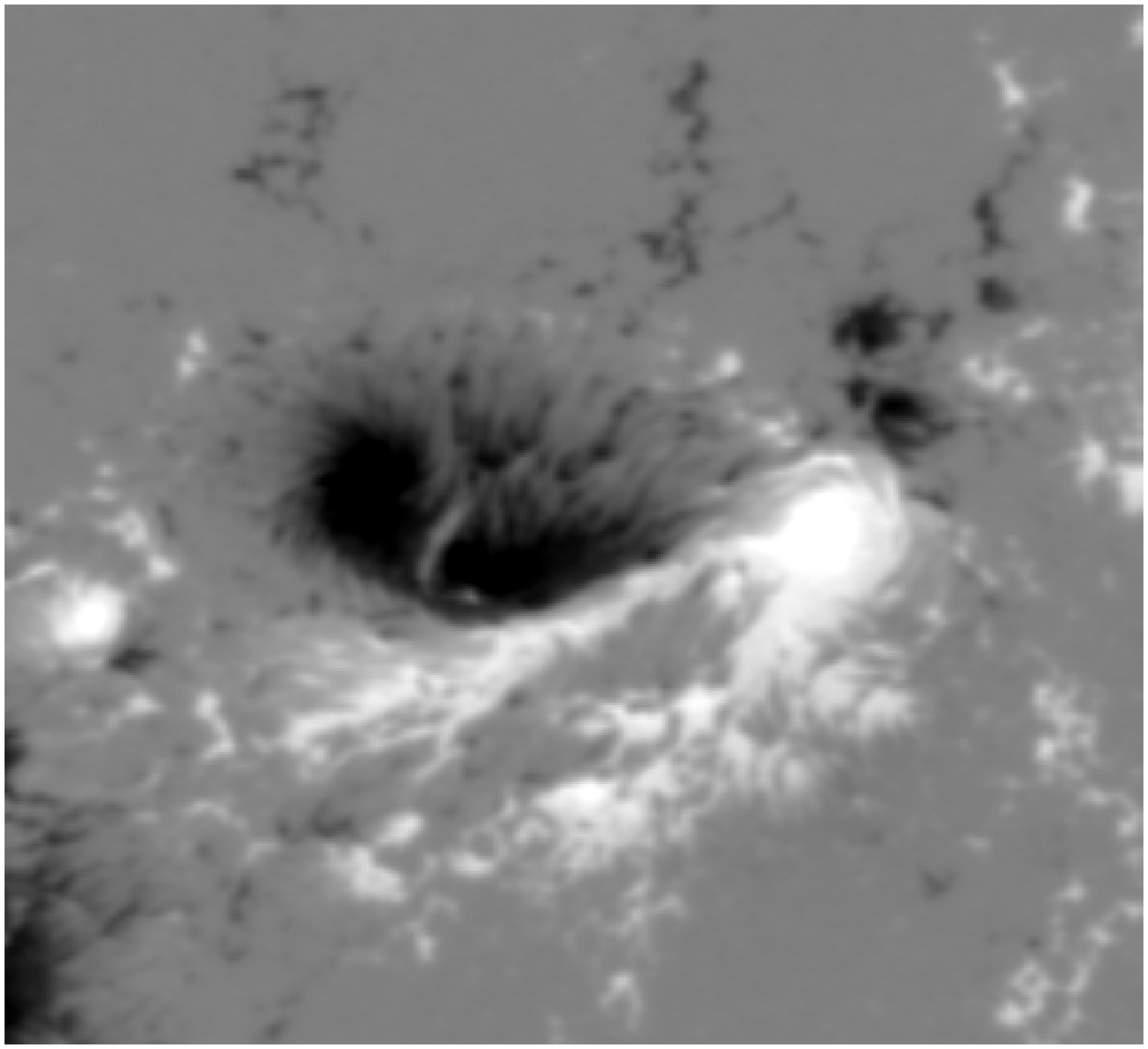}
\end{center}
\vspace*{-7mm}
\caption{A frame of a HMI magnetogram movie showing the dynamics of the flare ribbons (\url{http://sun.stanford.edu/~sasha/X2_flare_HMI/hmi_mag_transient_X2_flare.mpg}). White and black colors show magnetic field of positive and negative polarity in the central sunspot of AR 1158. The color map is saturated at 1200 G.}\label{fig2}
\begin{center}
\includegraphics[width=0.7\textwidth]{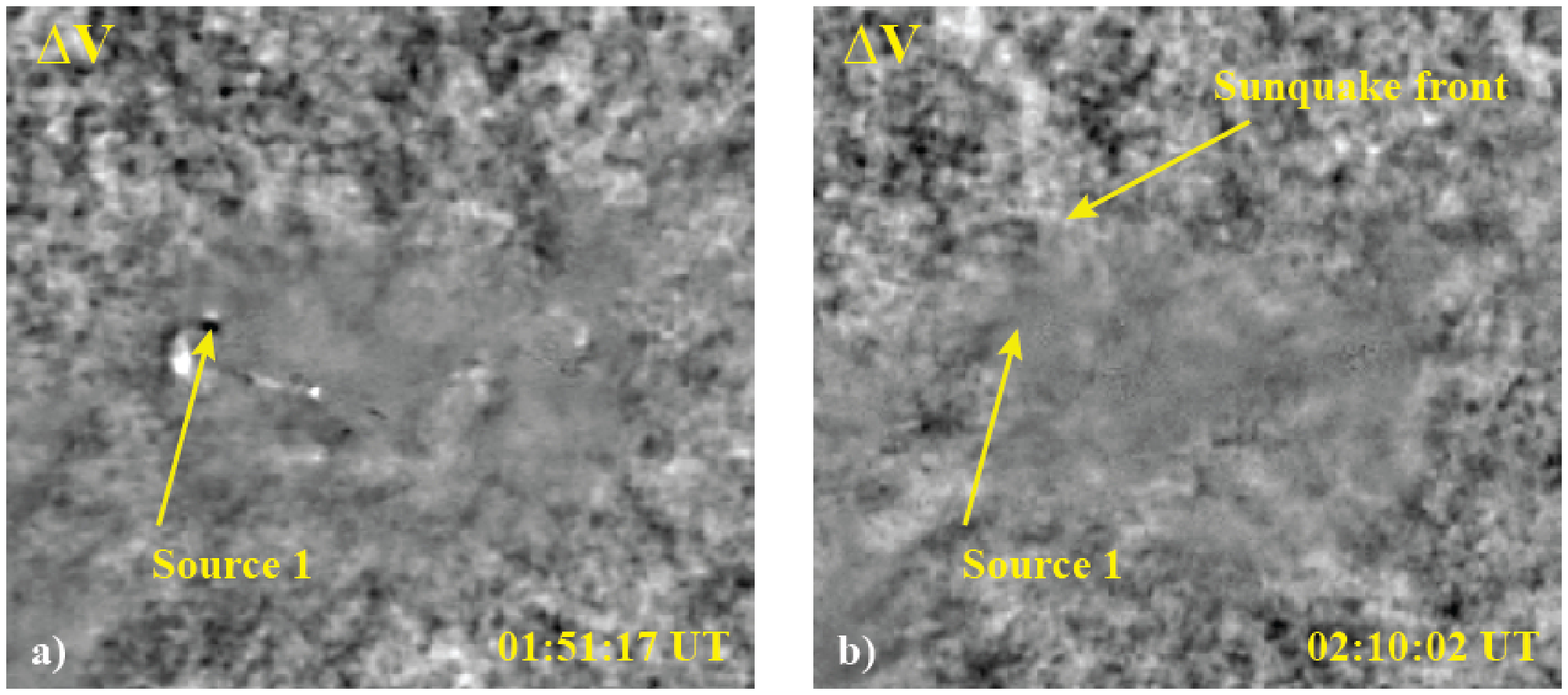}
\end{center}
\vspace*{-7mm}
\caption{Doppler velocity differences
 at the moments of: a) the flare impact at 01:51:17 UT, and b) about 19 min later at 02:10:02 UT, showing the sunquake wavefront. These are two frames of the running difference movie (\url{http://sun.stanford.edu/~sasha/X2_flare_HMI/hmi_sunquake_X2_flare_Feb_15_2011.mpg}).}\label{fig3}
\end{figure}

The X2.2 flare of February 15, 2011, occurred in the central sunspot of active region NOAA 1158, which had a $\delta$-type magnetic configuration (Fig.~\ref{fig1}). According to the GOES-15 soft X-ray measurements, the flare started at 01:44, reached maximum at 01:56 and ended at 02:06~UT. The flare signals are clearly detected in all HMI observables, and show that the flare had a typical two-ribbon structure with the ribbons located on both sides of the magnetic neutral line. This is  well seen in the magnetogram movie (Fig.~\ref{fig2}).
\begin{figure}[t]
\begin{center}
\includegraphics[width=0.55\textwidth]{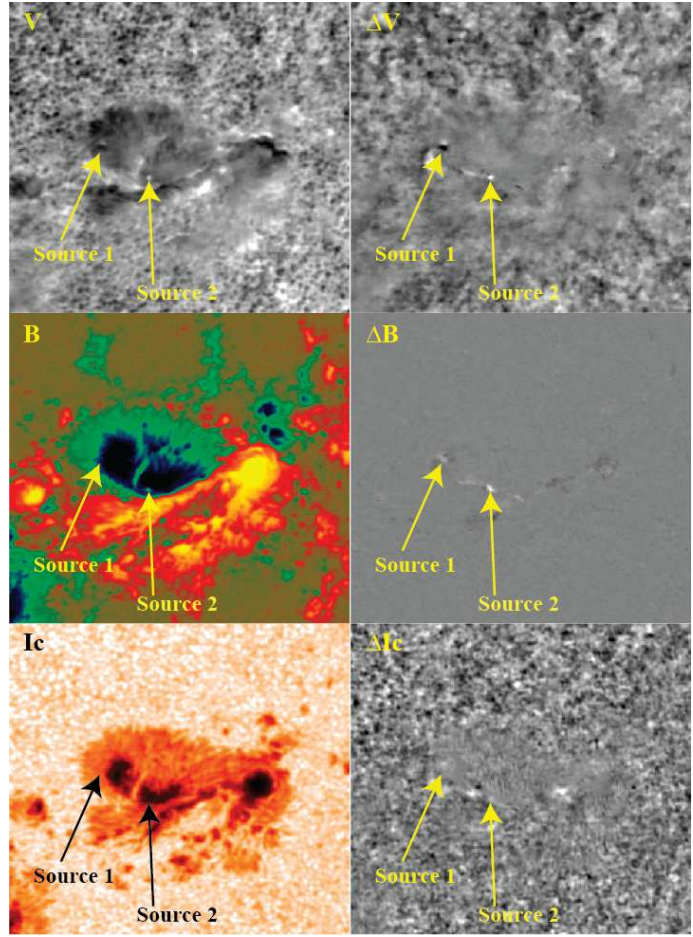}
\end{center}
\vspace*{-5mm}
\caption{Images in left panels show the Doppler velocity, $V$, line-of-sight magnetic field, $B$, and continuum intensity, $I_c$, taken 2011.02.15\_01:54:39\_TAI. The right panels show the differences between these images and the corresponding images taken 45 sec earlier.}\label{fig4}
\end{figure}

The sunquake event is revealed in the running difference movie of the Doppler velocity data (Fig.~\ref{fig3}). The sunquake wave appears in the upper left part about 20 minutes after the initial flare impact visible as a two-ribbon structure. The wave has an elliptical shape. The wavefront  traveling outside the magnetic region in the North-East direction is most clearly visible. In the opposite direction the wave travels through the magnetic field of sunspots, and its amplitude is significantly suppressed. This is a general property of acoustic waves on the Sun.

\begin{figure}[t]
\begin{center}
\includegraphics[width=0.9\textwidth]{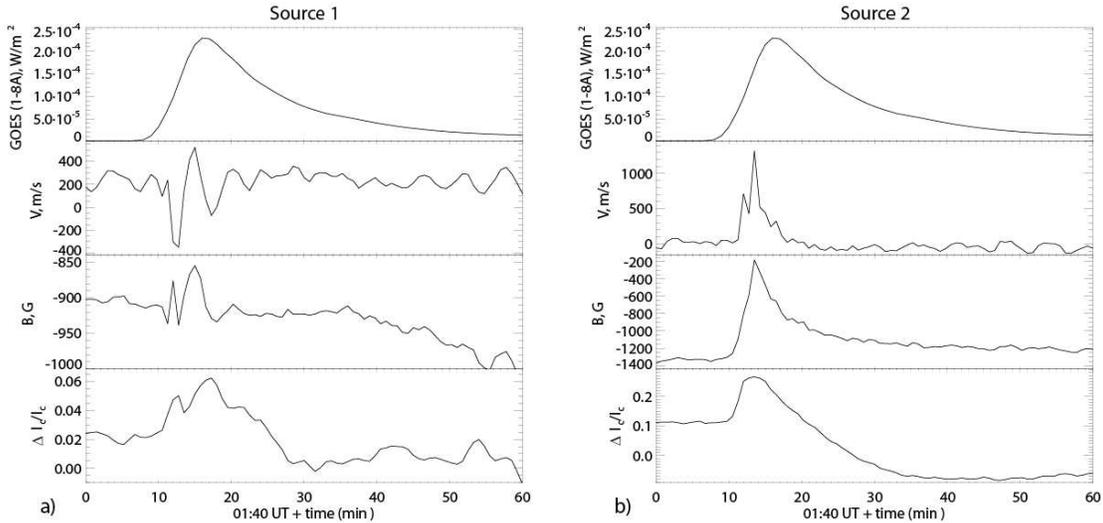}
\end{center}
\vspace*{-5mm}
\caption{Variations of the soft X-ray flux from GOES-15, Doppler velocity, magnetic field and continuum intensity at: a) Source 1; b) Source 2.}\label{fig5}
\end{figure}

The flare ribbons consist of individual patches representing impacts of flare impulses. In these data it is easy to find that the location of the sunquake source was in the penumbra area near the edge of the active region. These area is identified as `Source 1' in Figures~\ref{fig1},~\ref{fig3} and \ref{fig4}. It is characterized by strong and rapid variations of the Doppler velocity and magnetic field, and by an impulsive increase of the continuum intensity (Fig.~\ref{fig5}a).

The sunquake source is associated with one of the impacts located along the flare ribbons. There were strong photospheric impacts in several other locations. However, these impacts did not provide clearly visible seismic waves. The reason for this is not clear. For a comparison, in Figure~\ref{fig5}b we show the properties of one of the strongest compact sources identified as `Source 2'. This source was located in a region of strong magnetic field in the sunspot outer umbra near the magnetic neutral line. During the impact, the HMI data show a strong increase of the Doppler velocity, indicating downflows, a sharp impulsive decrease of the magnetic field strength, which relaxed to to a value lower the the pre-flare strength, and an increase in the continuum intensity brightness. All the variations in Source 2 are stronger than in Source 1, but it did not generate strong sunquake ripples.
 It seems that the main difference between these two places of the flare impact onto the photosphere of the Sun is that Source 1 was located in a region of relatively weak magnetic field contrary to Source 2, which was in strong field. In addition, Source 1 was more variable and moving. It started at the outer boundary of the penumbra (bright point at the Source 1 arrow in Fig.~\ref{fig3}a) and then moved into the penumbra, generating a localized wave-like motion in this part of the penumbra. Perhaps, the dynamic nature of the flare impact is important for understanding the mechanism of sunquakes. The strong magnetic field in Source 2 probably restricts wave motion.

\section{Discussion}

The first observations of the sunquake event from SDO/HMI revealed very interesting properties of the flare impact onto the solar photosphere. The HMI data with the significantly higher resolution  than the previous SOHO/MDI observations provide a new insight into the dynamics of the flare impact and the sunquake source. The preliminary analysis indicates that seismic flare waves are generated by the impact in the region of a relatively weak magnetic field of the sunspot penumbra. It is curious that a significantly stronger impact in a region of high magnetic field strength did not provide seismic waves of a comparable magnitude.

A characteristic feature of this sunquake is
anisotropy of the wave front: the observed wave amplitude is much
stronger in one direction than in the others. This was observed also in
previous events. In particular, the
seismic waves excited during the October 28, 2003, flare had the
greatest amplitude in the direction of the expanding flare ribbons.
The wave anisotropy was attributed to the moving source of the
hydrodynamic impact, which is located in the flare ribbons
\citep{Kosovichev2006c,Kosovichev2006b}. The motion of flare
ribbons is often interpreted as a result of the magnetic
reconnection processes in the corona. When the reconnection region
moves up it involves higher magnetic loops, the footpoints of which
are further apart. This may explain the expanding flare ribbons (as places
of the photospheric flare impacts) and
the association of sunquakes with the ribbon sources.
In this event, the sunquake had a similar dynamical property:
it started at an outer boundary of the sunspot penumbra and then quickly moved
in the penumbra region. This was accompanied by a wave-like motion of this part
of the penumbra. This is certainly an interesting phenomenon, which requires
further investigation.
Of course, there might be other reasons for the
anisotropy of the wave front, such as inhomogeneities in
temperature, magnetic field and plasma flows. However, the source
motion seems to be quite important for generating sunquakes.


\end{document}